# Effects of Transverse Force on Dusty Fluid Flow over a Linear Stretching Sheet

Subhrajit Kanungo[1], Pradeep Kumar Tripathy[2], Tumbanath Samantara[3,*]
[1]Gandhi Institute For Technology, Bhubaneswar, Odisha, India
[2]Department of Mathematics & Science, UtkalGouravMadhusudan Institute of Technology, Rayagada, Odisha, India
[3]Centurion University of Technology and Management, Odisha, India

[1]banty96kanungo@gmail.com,
[2]tripathypk2@gmail.com, [3]tnsamantara@gmail.com

**Abstract**

This research presents a numerical investigation of the flow and heat transfer of a steady dusty flow over a linear horizontal stretching sheet. Transverse force effects have been taken into account. The flow problem's formulation comprises of highly nonlinear PDEs that have been transformed into systems of ODEs by using similarity transformation. Then the ODEs has been solved numerically by using Shooting technique followed by RungeKutta 4th order method that is incorporated in BVP4C tool of the MATLAB software. The effects various flow parameters, such as the Prandtl number, Eckert number and transverse force on the flow geometry has been investigated. The overall findings are displayed in graphs and tables, and it is discovered that the transverse force reduces the velocity of the particle phase in the flow. Our results has been validated with existing literatures and found agrees with very good way.

**Introduction**

The influence of a transversal force across a stretching sheet and the continuous two-phase flow of a fluid have a stronger impact on engineering and manufacturing processes and industries. The compressor, a device that uses work or power, was designed using the principles of constant two-phase flow. Stretching sheets is used in a variety of industrial processes, including manufacturing paper, plastic sheets, coloring sheets, and more. Many academics have expressed a special interest in working on flow and heat transfer of fluid particle suspension across stretchy sheet with the influence of transvers force because of the vast number of applications.

Soo [1] has formulated the governing Equations for the two-phase flow and they analyzed the case of laminar boundary layer flow. Study of transvers force has a greater impact on the flow of fluid as well as particle. An approximation expression for the shear lift force on a spherical particle at an infinite Reynolds number has been studied by R.

Mei[2]. Saffmann [3] has shown that the particle experience a transvers force due to the combination of slip-shear and lift force. Otterman [4] has deliberate a study on laminar mixing of dusty fluid with clear fluid and has analyzed the effect of transvers force on the flow fluid. Marbel [5] solved the boundary layer flow Equation by series expansion method, which is most crucial for the down-stream region of the plate. Singleton [6] also extended this study to compressible gas particle flow. Datta and Mishra [7] investigated the boundary layer flow of a dusty fluid dusty fluid over a semi-infinite flat plate. Jain and Ghosh [8] also analyzed the particle momentum Equation in the transvers direction and considered the effect of transvers force on the flow field in addition to drag force. Prabha and Jain [9] have studied the gas particulate boundary layer flow with no pressure gradient. The first ever exploration of boundary layer flow on a stretching sheet surface was explored by Sakiadis[10-11]. The problem of the boundary layer flow which has been produced by a linear stretching sheet has been studied by Crane. [12] Then so many academics and scholars stretched their research work on linear stretching sheet in view of different physical parameters. "A detailed study about the heat and mass transfer over a linear stretching sheet has been analyzed by Gupta. Tripathy [13,14,15,19] has studied on the two phase boundary layer flow and heat transfer with non-uniform grid. Samantara T.N et.al. [16,17,18] have studied impact of electrification of particles in flow geometry of a horizontal plate, inclined stretching sheet and jet flow."

Hence in this study an investigation on the flow and heat transfer of a fluid over a linear horizontal stretching sheet with effect of transvers force has been carried out.

**Mathematical modelling and solution of the problem**

The steady boundary layer flow of fluid over a linear stretching sheet is taken in Figure 6.1. The wall is stretched linearly wit velocity $U_w(x) = cx$ due to the application of two interacting opposite force applied on the wall. The X-axis is considered as the flow of the fluid and the Y-axis is normal to it. Initially the both fluid and the particle are in rest position.

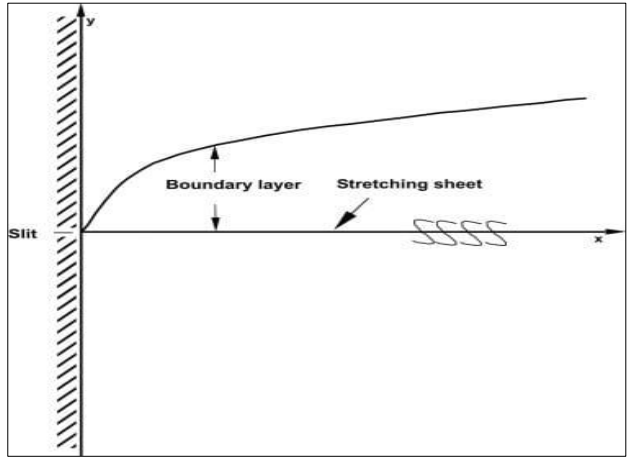

**Figure 1 Geometry of the flow problem**

Despite the above consideration of the boundary layer, the governing Equation of continuity, momentum and energy Equation for both fluid and the particle phase with appropriate notation are

$$\frac{\partial u}{\partial x} + \frac{\partial v}{\partial y} = 0 \tag{1}$$

$$\frac{\partial}{\partial x}(\rho_p u_p) + \frac{\partial}{\partial y}(\rho_p v_p) = 0 \tag{2}$$

$$(1-\varphi)\rho\left(u\frac{\partial u}{\partial x} + v\frac{\partial u}{\partial y}\right) = (1-\varphi)\mu\frac{\partial^2 u}{\partial y^2} - \frac{1}{\tau_p}\varphi\rho_s(u-u_p) \tag{3}$$

$$\varphi\rho_s\left(u_p\frac{\partial u_p}{\partial x} + v_p\frac{\partial u_p}{\partial y}\right) = \frac{\partial}{\partial y}\left(\varphi\mu_s\frac{\partial u_p}{\partial y}\right) + \frac{1}{\tau_p}\varphi\rho_s(u-u_p) + \varphi\frac{0.73(\rho/\rho_s)^{1/2}}{\tau_p^{1/2}}\left|\frac{\partial u}{\partial y}\right|^{1/2}(u-u_p) \tag{4}$$

$$\varphi\rho_s\left(u_p\frac{\partial v_p}{\partial x} + v_p\frac{\partial v_p}{\partial y}\right) = \frac{\partial}{\partial y}\left(\varphi\mu_s\frac{\partial v_p}{\partial y}\right) + \frac{1}{\tau_p}\varphi\rho_s(v-v_p) \tag{5}$$

$$(1-\varphi)\rho c_p\left(u\frac{\partial T}{\partial x} + v\frac{\partial T}{\partial y}\right) = (1-\varphi)k\frac{\partial^2 T}{\partial y^2} + \frac{1}{\tau_T}\varphi\rho_s c_s(T_p - T)$$
$$+ \frac{1}{\tau_p}\varphi\rho_s(u-u_p)^2 + (1-\varphi)\mu\left(\frac{\partial u}{\partial y}\right)^2 \tag{6}$$

$$\varphi\rho_s c_s\left(u_p\frac{\partial T_p}{\partial x} + v_p\frac{\partial T_p}{\partial y}\right) = \frac{\partial}{\partial y}\left(\varphi k_s\frac{\partial T_p}{\partial y}\right) - \frac{1}{\tau_T}\varphi\rho_s c_s(T_p - T)$$
$$- \frac{1}{\tau_p}\varphi\rho_s(u-u_p)^2 + \varphi\mu_s\left[u_p\frac{\partial^2 u_p}{\partial y^2} + \left(\frac{\partial u_p}{\partial y}\right)^2\right] \tag{7}$$

With boundary condition

$$u = U_w(x), v = 0, T = T_w = T_\infty + A\left(\frac{x}{l}\right)^2; \qquad \text{as } y \to 0$$

$$u = 0, u_p = 0, v_p = v, T = T_p = T_\infty; \qquad \text{as } y \to \infty \quad (8)$$

Where $U_w(c) = cx$, is the linear stretching sheet velocity, $c$ be the stretching rate positive constant. $T_w$ be the wall temperature and $A$ is the positive constant. $l = \sqrt{\frac{v}{c}}$ is the characteristic length. For gas Equation $\tau_p = \tau_T$ if $\frac{c_s}{c_p} = \frac{2}{3Pr}$ and $k_s = k\frac{c_s}{c_p}\frac{\mu_s}{\mu}$.

The Equation (1) is identical satisfied the function $\psi(x, y) = \sqrt{cv}xf(\eta)$ such that $u = \frac{\partial \psi}{\partial y}$, $v = -\frac{\partial \psi}{\partial x}$.

We further lead the following transformations in the Equations (2) to (7), to convert the governing equations into a set of Ordinary Differential Equations,

$$u = cxf'(\eta), \qquad v = -\sqrt{cv}f(\eta), \qquad \eta = \sqrt{c/v}\,y$$

$$u_p = cxF(\eta), \qquad v_p = \sqrt{cv}G(\eta), \qquad \rho_r = \rho_p/\rho = H(\eta)$$

$$\theta(\eta) = \frac{T - T_\infty}{T_w - T_\infty}, \quad \theta_p(\eta) = \frac{T_p - T_\infty}{T_w - T_\infty} \qquad (9)$$

Where $T - T_\infty = A\left(\frac{x}{l}\right)^2 \theta(\eta), \quad T_p - T_\infty = A\left(\frac{x}{l}\right)^2 \theta_p(\eta)$

We get the a couple of non-linear ordinary differential Equation as

$$H' = -\frac{(HF + HG')}{G} \qquad (10)$$

$$f''' = f'^2 - ff'' + \frac{1}{1-\varphi}\beta H(f' - F) \qquad (11)$$

$$F'' = \frac{1}{\epsilon}[F^2 + F'G - \beta(f' - F)] \qquad (12)$$

$$G'' = \frac{1}{\epsilon}\left[GG' + \beta(f + G) - 0.73\sqrt{\beta}F_t\sqrt{f''}(f' - F)\right] \qquad (13)$$

$$\theta'' = (2f'\theta - f\theta')Pr - \frac{2}{3}\frac{H\beta}{(1-\phi)}(\theta_p - \theta) - \frac{H\beta PrEc}{(1-\phi)}(f' - F)^2 - PrEcf''^2 \qquad (14)$$

$$\theta_p'' = \frac{\epsilon}{Pr}\left[\begin{array}{c}2F\theta_p + G\theta_p' + \beta(\theta_p - \theta) + \frac{3}{2}PrEc\beta(f' - F)^2 \\ -\frac{3}{2}\epsilon PrEc(FF'' + F'^2)\end{array}\right] \qquad (15)$$

Non-Dimensional Boundary Condition:

$G'(\eta) = 0, f(\eta) = 0, f'(\eta) = 1, F'(\eta) = 0, \theta(\eta) = 1, \theta'_p(\eta) = 0; as \eta \to 0$

$f'(\eta) = 0, F(\eta) = 0, G(\eta) = -f(\eta), H(\eta) = \omega, \theta(\eta) \to 0, \theta_p(\eta) \to 0; as \eta \to \infty$

(16)

**Numerical Methods and Validations**

The highly non-linear ordinary differential Equation (10) to (15) with boundary condition (16) are computed numerically by using Runge-Kutta 4$^{th}$ order scheme with bvp4c method of Matlab software which is more truthful and well verified process. We equate our surface temperature $-\theta'(0)$ result for some different value of Prandtl number with the results of some previous researchers which are shown in Table 1.

**Table 1 result validating table**

| Prandtl number, Pr | Ishaket. al.[20] | Subhaset.al. [21] | Giresshaet. al.[22] | Chen [23] | Gurbkaet. al. [24] | Mukhop-adhyaet.al [25] | Current Study |
|---|---|---|---|---|---|---|---|
| 0.72 | ---------- | 1.0885 | 1.0885 | 1.0885 | 1.0885 | 1.0885 | 1.0884 |
| 1.0 | 1.3333 | 1.3333 | 1.3333 | 1.3333 | 1.3333 | 1.3333 | 1.3333 |
| 3.0 | 2.5097 | --------- | 2.5097 | 2.5097 | ---------- | 2.5097 | 2.5097 |
| 10.0 | 4.7969 | 4.7969 | 4.7969 | 4.7969 | 4.7969 | ---------- | 4.7969 |

From above table it is remarked that our results are mostly accurate with previous researchers and hence our problem is validate.

**Results and Discussion**

Figure 2 and Figure 3 depicts the effects of Prandtl number on heat transfer profile of flow field. It is concluded that heat transfer decreases with rise of Prandtl number in both the phases.

Figure 4-7 depicts the effects ofparticle interaction parameter $\beta$ on flow and heat transfer profile of flow field. It is witnessed that particle interaction parameter $\beta$ has negligible effects in flow and heat transfer of fluid phase but has significant effects on particle phase of flow and heat transfer profile. The flow and heat transfer increases with increase of interaction parameter $\beta$.

Figure 8-9 depicts the effects of Eckert number (Ec) on heat transfer profile of flow field. From the graph it is observed that the heat transfer rises with increasing the value of Ec in both fluid and particle phases.

Figure 10-13 depicts the effect of diffusion parameter ($\epsilon$) on velocity and heat transfer profileof the flow geometry. It is observed from the figure that the effect of diffusion parameter has negligible effect in fluid phase but have remarkable effect in case of particle phase. The velocity and heat transfer of particle phase increases with increasing the value of diffusion parameter.

Figure 14-17 depicts the effect of transverse force on flow and heat transfer of flow geometry. It is detected from the graph that transverse force has very negligible effect on flow and heat transfer in case of fluid phase but has some effects on flow and heat transfer in particle phase. The velocity and heat transfer decreases as the transverse force value increases. This happens due to acting of transverse force perpendicular to the flow direction.

The effect of different parameters on skin friction and andNusselt number is represented in following Table 6.1. From the table it is observed that increasing the transverse force causes the lowering of Skin friction and Nusselt number.

## 6.5 CONCLUSION

The effects of different parameters on flow over Linear Stretching Sheet has been studied and conclusions are as follows. The transverse force which is applied to flow field is perpendicular to the flow and hence hinders the velocity and heat transfer of the flow following the skin friction and nusselt number.

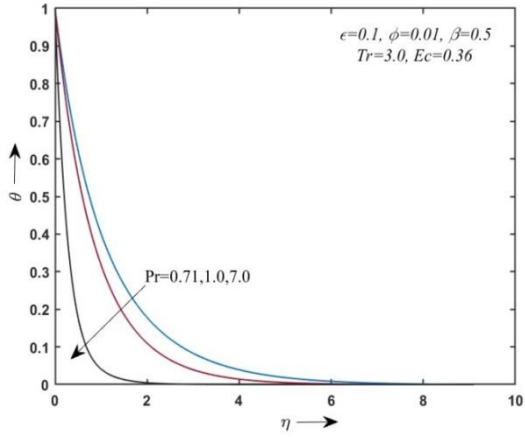

**Figure 2** Effect of Prandtl number on fluid temperature

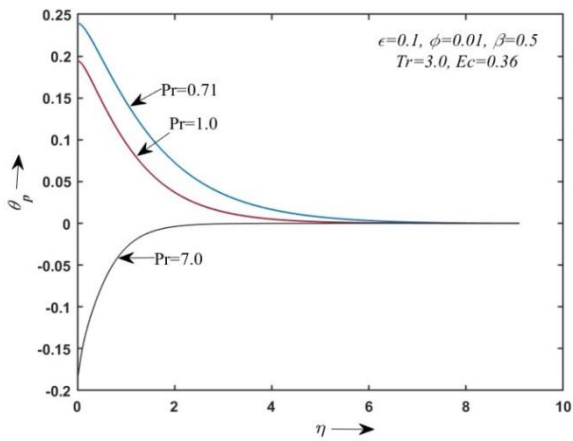

**Figure 3** Effect of Prandtl number on particle temperature

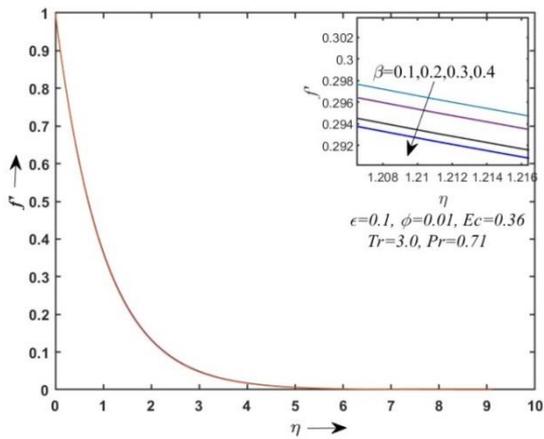

**Figure 4** Effect of particle interaction parameter $\beta$ on fluid velocity

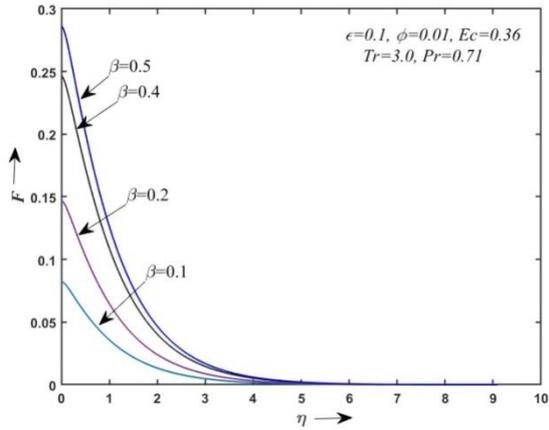

**Figure 5 Effect of particle interaction parameter $\beta$ on particle velocity**

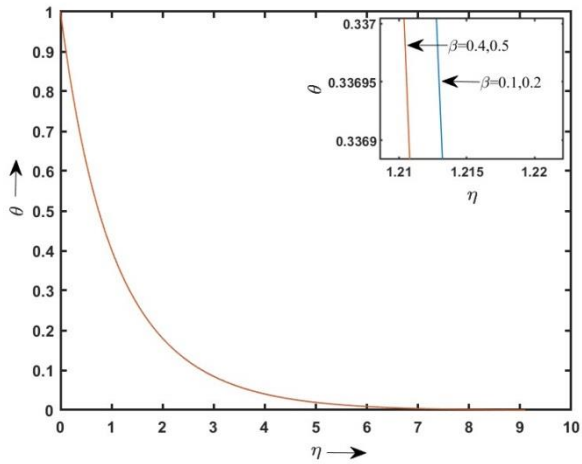

**Figure 6 Effect of particle interaction parameter $\beta$ on fluid temperature**

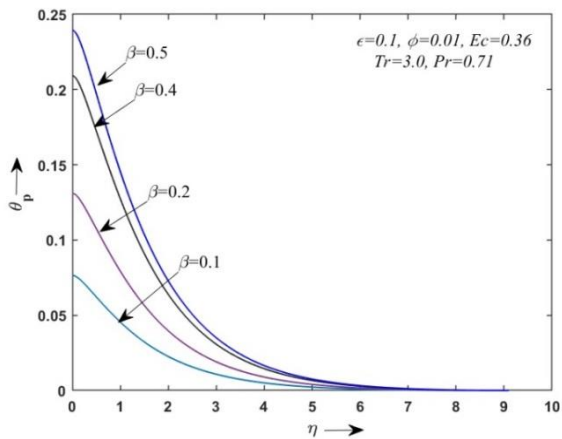

**Figure 7 Effect of particle interaction parameter $\beta$ on particle temperature**

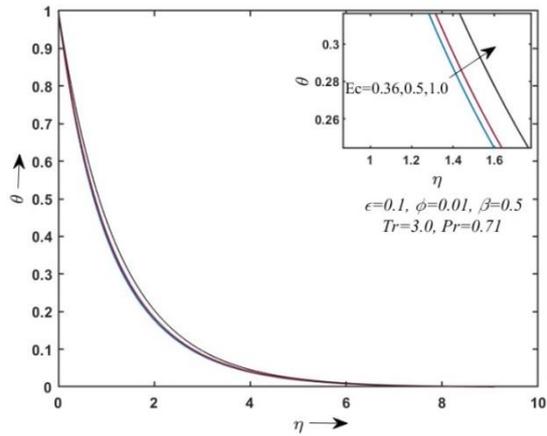

**Figure 8 Effect of Eckert number (Ec) on fluid temperature**

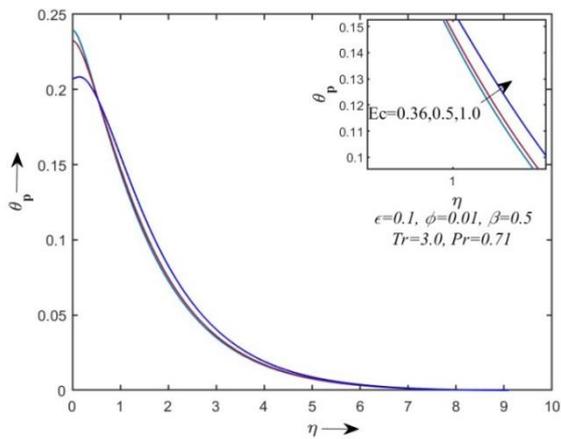

**Figure 9 Effect of Eckert number (Ec) on particle temperature**

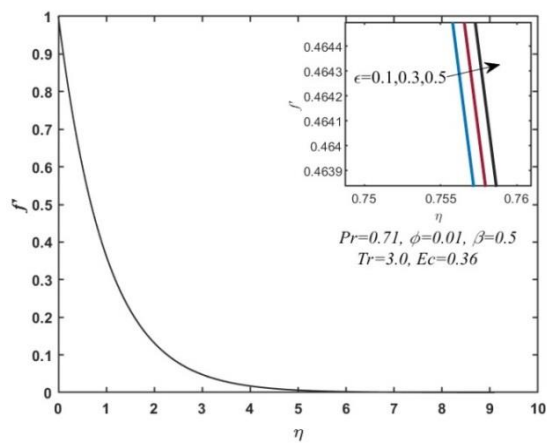

**Figure 10 Effect of diffusion parameter ($\epsilon$) on fluid velocity**

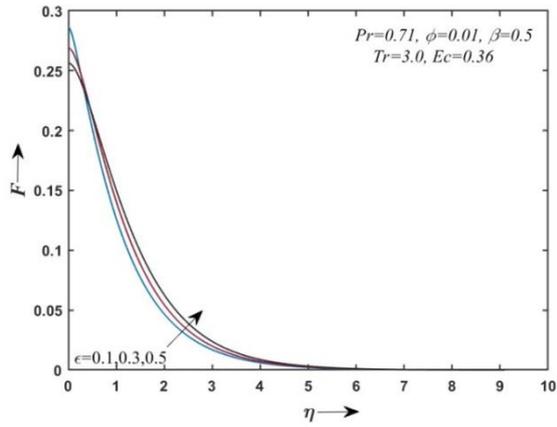

**Figure 11 Effect of diffusion parameter ($\epsilon$) on particle velocity**

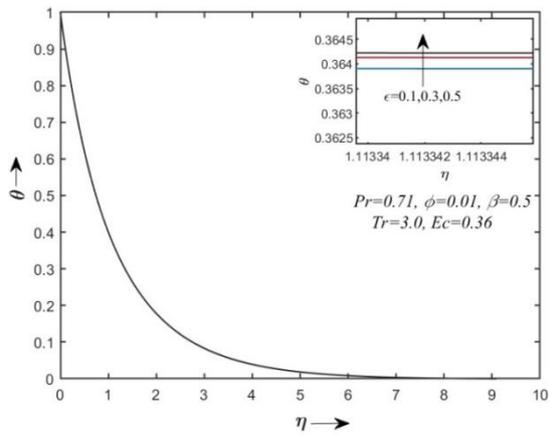

**Figure 12 Effect of diffusion parameter ($\epsilon$) on fluid temperature**

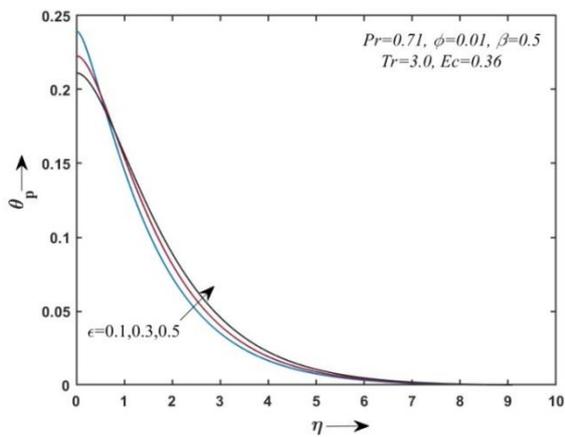

**Figure 13 Effect of diffusion parameter ($\epsilon$) on particle temperature**

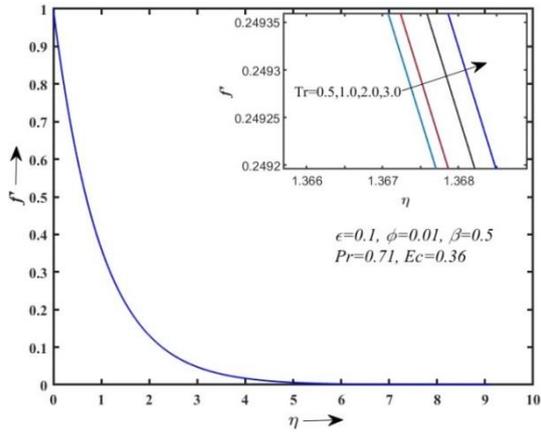

**Figure 14** Effect of transvers force ($Tr$) on fluid velocity

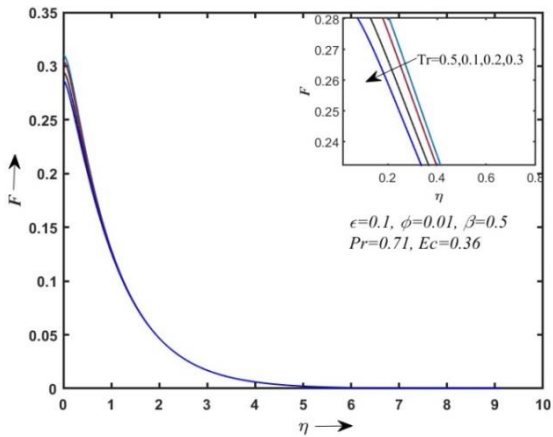

**Figure 15** Effect of transvers force ($Tr$) on particle velocity

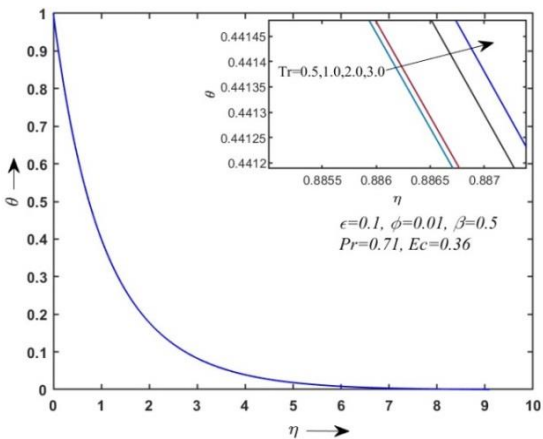

**Figure 16** Effect of transvers force ($Tr$) on fluid temperature

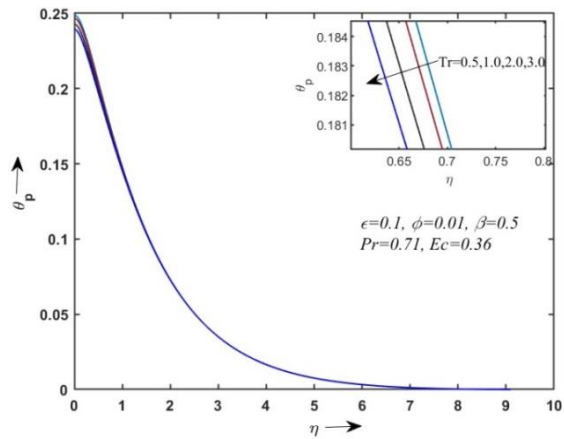

**Figure 17 Effect of transvers force $(Tr)$ on particle temperature**

**Table 2 Effect of $Pr, Ec, Tr, \beta, \epsilon$ on skin friction and Nusselt number**

| $Pr$ | $Ec$ | $Tr$ | $\beta$ | $\epsilon$ | Skin Friction $f''(0)$ | Nusselt Number $-\theta'(0)$ |
|---|---|---|---|---|---|---|
| 0.71 | 0.36 | 3.0 | 0.5 | 0.1 | 1.34618 | 1.25240 |
| 1.0 | | | | | 1.34618 | 1.34728 |
| 7.0 | | | | | 1.34618 | 2.80682 |
| 0.71 | 0.36 | 3.0 | 0.5 | 0.1 | 1.34618 | 1.25240 |
| | 0.5 | | | | 1.34618 | 1.19462 |
| | 1.0 | | | | 1.34618 | 0.98825 |
| 0.71 | 0.36 | 3.0 | 0.5 | 0.1 | 1.34618 | 1.25240 |
| | | 0.5 | | | 1.34889 | 1.25399 |
| | | 1.0 | | | 1.34797 | 1.25343 |
| | | 2.0 | | | 1.34686 | 1.25278 |
| 0.71 | 0.36 | 3.0 | 0.5 | 0.1 | 1.34618 | 1.25240 |
| | | | 0.1 | | 1.34221 | 1.25051 |
| | | | 0.2 | | 1.34339 | 1.25103 |
| | | | 0.4 | | 1.34534 | 1.25196 |
| 0.71 | 0.36 | 3.0 | 0.5 | 0.1 | 1.34618 | 1.25240 |
| | | | | 0.3 | 1.34644 | 1.25260 |
| | | | | 0.5 | 1.34696 | 1.25282 |